# PupiNet: Seamless OCT-OCTA Interconversion Through Wavelet-Driven and Multi-Scale Attention Mechanisms

Renzhi Tian
School of Computer Science and Technology
University of Chinese Academy of Sciences
Beijing, China
799842024@qq.com

Jinjie Wang
School of Computer Science and Technology
University of Chinese Academy of Sciences
Beijing, China
3189425293@qq.com

Wei Yang
School of Computer Science and Technology
University of Chinese Academy of Sciences
Beijing, China
sheepwei_2023@qq.com

Weizhen Li
School of Computer Science and Technology
University of Chinese Academy of Sciences
Beijing, China
19100750506@163.com

Haoran Chen
School of Computer Science and Technology
University of Chinese Academy of Sciences
Beijing, China
202308611@stu.sicau.edu.cn

Yiran Zhu
School of Computer Science and Technology
University of Chinese Academy of Sciences
Beijing, China
ciaran_study@yeah.net

Chengchang Pan*
School of Computer Science and Technology
University of Chinese Academy of Sciences
Beijing, China
chpan.infante@qq.com

Honggang Qi*
School of Computer Science and Technology
University of Chinese Academy of Sciences
Beijing, China
hgqi@ucas.ac.cn

## ABSTRACT

Optical Coherence Tomography (OCT) and Optical Coherence Tomography Angiography (OCTA) are key diagnostic tools for clinical evaluation and management of retinal diseases. Compared to traditional OCT, OCTA provides richer microvascular information, but its acquisition requires specialized sensors and high-cost equipment, creating significant challenges for the clinical deployment of hardware-dependent OCTA imaging methods. Given the technical complexity of OCTA image acquisition and potential mechanical artifacts, this study proposes a bidirectional image conversion framework called PupiNet, which accurately achieves bidirectional transformation between 3D OCT and 3D OCTA.

The generator module of this framework innovatively integrates wavelet transformation and multi-scale attention mechanisms, significantly enhancing image conversion quality. Meanwhile, an Adaptive Discriminator Augmentation (ADA) module has been incorporated into the discriminator to optimize model training stability and convergence efficiency. To ensure clinical accuracy of vascular structures in the converted images, we designed a Vessel Structure Matcher (VSM) supervision module, achieving precise matching of vascular morphology between generated images and target images. Additionally, the Hierarchical Feature Calibration (HFC) module further guarantees high consistency of texture details between generated images and target images across different depth levels.

To rigorously validate the clinical effectiveness of the proposed method, we conducted a comprehensive evaluation on a paired OCT-OCTA image dataset containing 300 eyes with various retinal pathologies. Experimental results demonstrate that PupiNet not only reliably achieves high-quality bidirectional transformation between the two modalities but also shows significant advantages in image fidelity, vessel structure preservation, and clinical usability.

# 1 Introduction

Retinal Optical Coherence Tomography (OCT) is a imaging modality that can non-invasively acquire high-resolution 3D images by measuring backscattered light echoes[1], and is widely used in the diagnosis of ophthalmic diseases such as age-related macular degeneration, diabetic retinopathy, and glaucoma [2-4]. OCTA is a novel imaging modality based on OCT that can non-invasively obtain blood flow information in retinal microvasculature without using any contrast agents, providing significant clinical value for early detection of lesions such as choroidal neovascularization in wet age-related macular degeneration[5-7]. However, acquiring high-quality OCTA images typically requires specialized hardware support and may introduce background noise and motion artifacts that reduce image quality, significantly increasing equipment costs and operational complexity.

In recent years, deep learning technologies have achieved breakthrough progress in medical image processing, demonstrating enormous potential particularly in cross-modal data conversion [6] Through deep generative models, researchers can now generate OCTA images from conventional OCT images, avoiding expensive hardware upgrades and complex operational procedures. Methods based on Generative Adversarial Networks (GANs) enable models to extract vascular information from OCT images and generate OCTA images with blood flow contrast. Furthermore, achieving reverse conversion from OCTA to OCT can provide more comprehensive reference data for clinical diagnosis.

This study proposes PupiNet, an efficient bidirectional transformation model for OCT and OCTA images. By integrating wavelet transformation and multi-scale attention mechanisms in the generator, and introducing an Adaptive Discriminator Augmentation (ADA) module in the discriminator, this model achieves the ability to generate high-quality OCTA and OCT images without relying on additional hardware. To ensure the clinical accuracy of generated images, we designed a Vessel Structure Matcher (VSM) supervision module to precisely control the consistency of vascular structures between generated images and target images. Meanwhile, the Hierarchical Feature Calibration (HFC) module further guarantees that the textures and details of generated images highly match the target images at different levels.

Through these technical innovations, PupiNet not only reduces medical equipment costs and operational complexity but also provides clinicians with more convenient and accurate diagnostic tools. We conducted rigorous experimental validation on a paired dataset of OCT and OCTA images from 300 eyes with various types of retinal diseases, and the results demonstrate that this method can stably generate high-quality medical images with significant clinical application value and promotion potential.

The contributions of this study are mainly reflected in the following three aspects:

- Innovative Deep Learning Model (PupiNet): We designed and implemented a deep learning model called PupiNet that efficiently accomplishes bidirectional transformation between OCT and OCTA. By integrating wavelet transformation and multi-scale attention mechanisms into the generator module, and introducing the Adaptive Discriminator Augmentation (ADA) module in the discriminator, we significantly improved the quality and accuracy of image conversion.
- Multi-level Structure Optimization: We developed a set of collaborative structure optimization techniques, including the Vessel Structure Matcher (VSM), Hierarchical Feature Calibration (HFC), and Adaptive Discriminator Augmentation (ADA). VSM ensures the consistency and accuracy of vascular structures between the converted images and the target images. HFC focuses on texture and detail calibration at different depth levels, ensuring high authenticity of generated images at various levels; ADA dynamically adjusts discriminator intensity based on loss, enabling the generator to more precisely capture subtle differences in data distribution.
- Clinical Application Value: This study not only achieved breakthroughs at the technical level but also deeply explored its potential for clinical applications. By reducing the need for expensive hardware, it lowered medical costs; by avoiding invasive procedures, it reduced patient discomfort; and by providing convenient and efficient diagnostic tools, it significantly improved clinical work efficiency. These advantages collectively promote early detection and treatment of ophthalmic diseases, carrying important clinical significance and social benefits.

# 2 Related Work

## 2.1 OCT and OCTA

Optical Coherence Tomography (OCT) is a non-invasive imaging technique that generates depth-resolved imaging through low-coherence interferometry, scanning the eye region to obtain

*Article Title Footnote needs to be captured as Title Note
†Author Footnote to be captured as Author Note

interference measurements and produce structural anatomical images. OCT primarily provides visualization of anatomical changes, but offers relatively low contrast between small blood vessels and tissues within retinal layers[8].

Optical Coherence Tomography Angiography (OCTA) utilizes the principle of diffraction particle motion of moving red blood cells to determine vascular locations in various parts of the eye without requiring intravascular dyes. It can visualize vessels down to the capillary level, providing detailed flow imaging of deep retinal vascular plexuses and the choroid[9]. However, OCTA application faces challenges such as background noise, artifacts, and differences between imaging protocols and devices, increasing equipment costs and operational complexity[10].

### 2.2 Generative Adversarial Networks

Generative Adversarial Networks (GANs)[11] have shown excellent performance in OCT to OCTA synthesis tasks. AdjacentGAN[12] uses three adjacent OCT-B scans as input to synthesize the middle OCTA-B scan image, enhancing continuity and spatiality. MultiGAN[13] uses OCT projection images to synthesize OCTA projection images between different retinal layers. TransPro[14] was the first to utilize 3D OCT data to synthesize OCTA data, effectively considering spatial coherence and overall structure, ensuring correlation between data.

This study builds upon previous work, further focusing on internal image associations, combining wavelet transformation with multi-scale attention mechanisms to effectively capture global information and local details of images. Meanwhile, it applies adaptive augmentation strategies to prevent the discriminator from prematurely stopping training, enhancing the robustness of the training process and improving the generator's capabilities.

## 3 Approach

The overall model is illustrated in Figure 1. In this study, we employed a multi-task deep learning model called PupiNet to perform OCT and OCTA conversion. Our model mainly consists of the following three components:

A ***3D pix2pix GAN*** integrating wavelet transformation, multi-scale attention, and adaptive discriminator augmentation as the backbone network;

A ***Vessel Structure Matcher*** (VSM) that ensures consistency and accuracy of vascular structures between converted images and target images;

A ***Hierarchical Feature Calibration*** (HFC) module that guarantees textures and details at specific levels in the generated images match those in the target images.

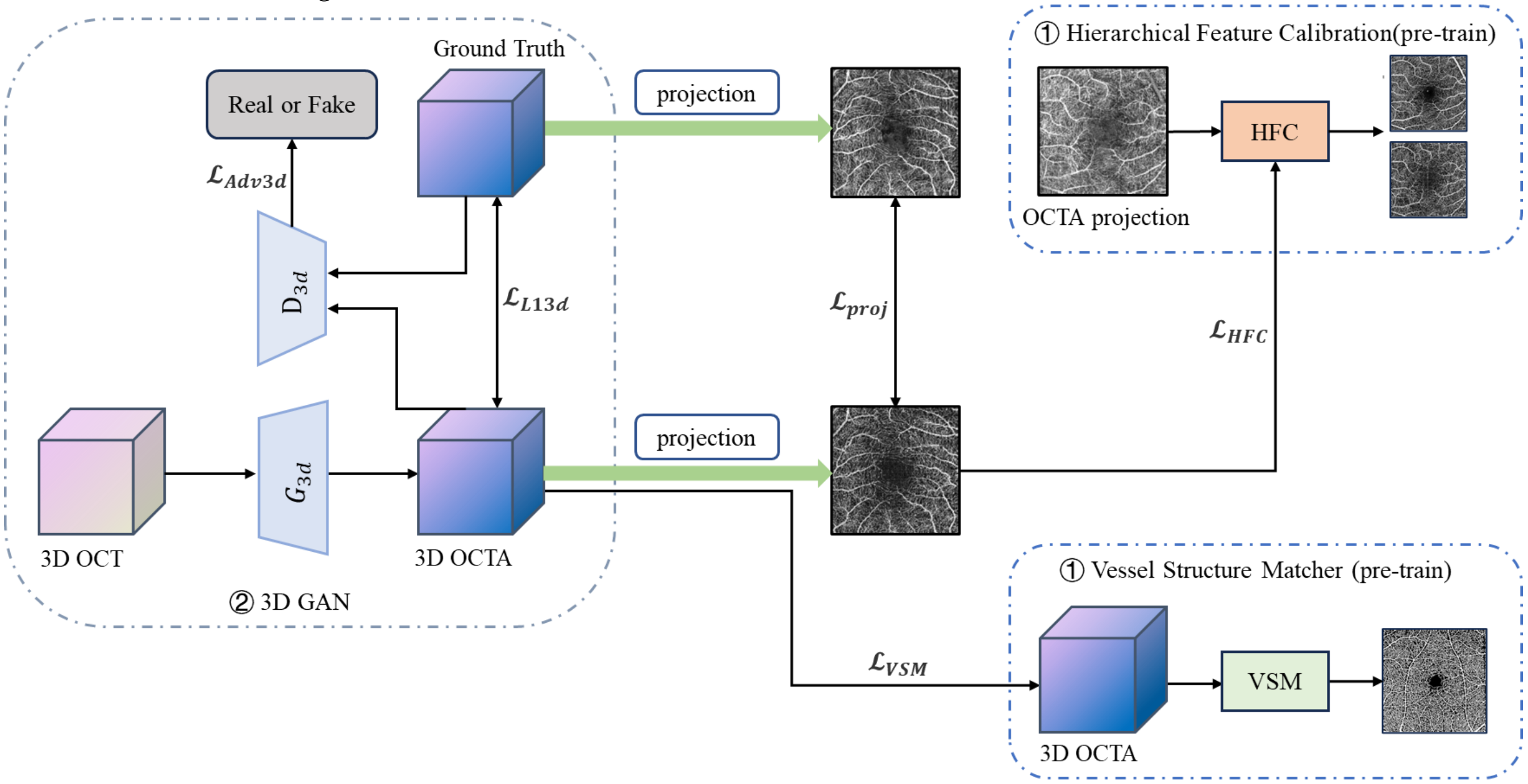

**Figure 1. The overall model architecture of our idea**

### 3.1 3D pix2pixGAN

Our proposed framework is based on the image conversion model pix2pix GAN[15], with key innovations and extensions. Unlike approaches that only process 2D B-scan images, we employ 3D convolutional neural networks to process complete volumetric data, overcoming spatial inconsistency issues caused by single-slice processing. This significantly enhances the three-dimensional consistency of converted images, laying the foundation for high-quality bidirectional transformation between 3D OCT and OCTA images. The generator architecture of 3D GAN is shown in Figure 2.

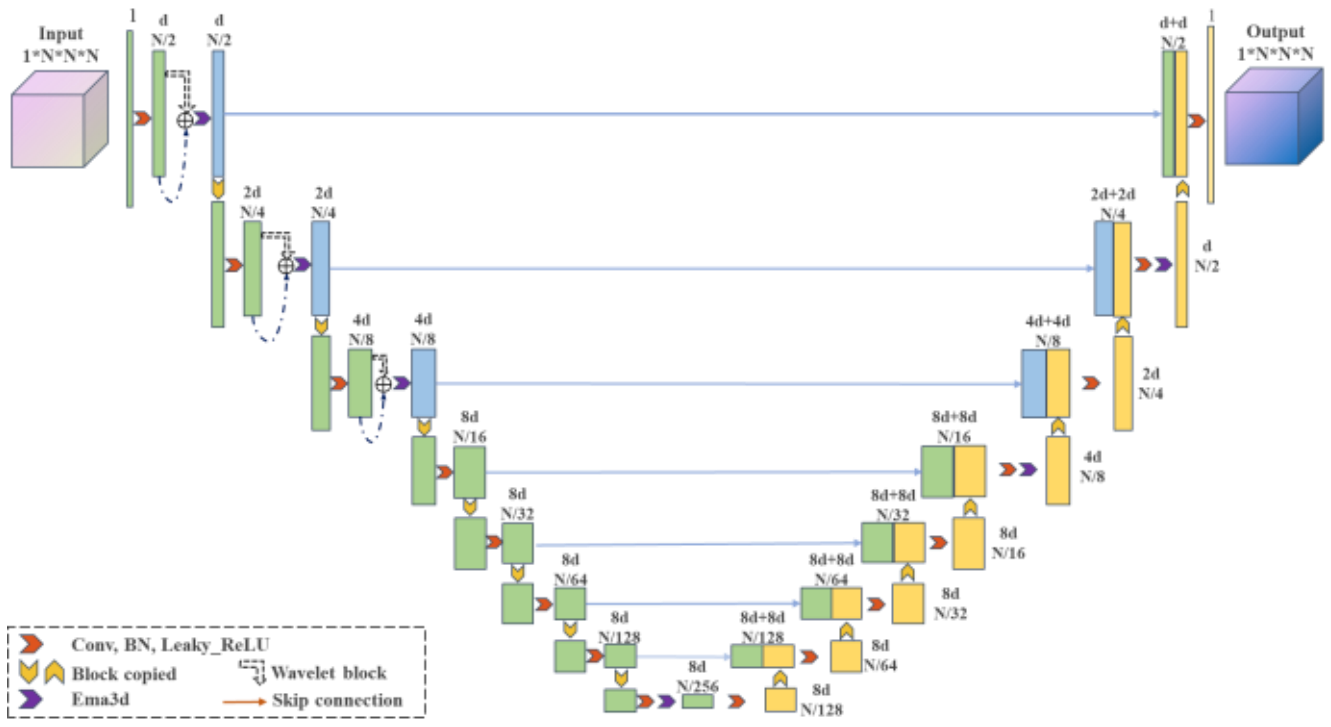

**Figure 2. The optimized generator part of the 3D GAN**

Additionally, to optimize model stability and generation quality, we integrated wavelet transformation[16] and multi-scale attention mechanisms[17] in the generator, and introduced Adaptive Discriminator Augmentation (ADA)[18] in the discriminator. For details on the structure of the wavelet transformation, please refer to Figure 3.

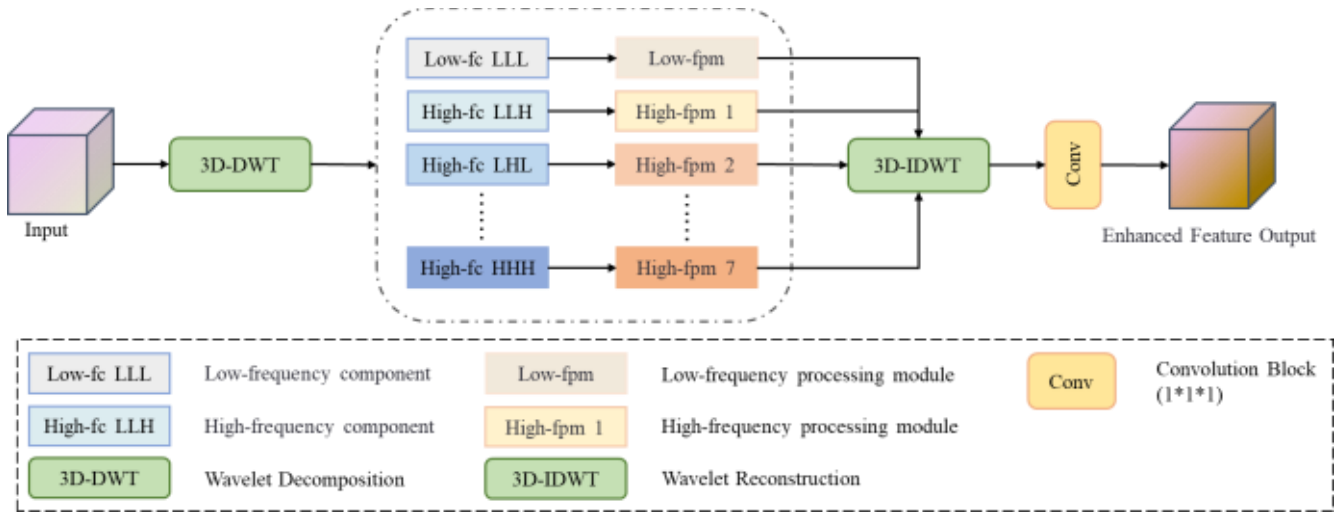

**Figure 3. The main process of wavelet transformation**

Wavelet transformation, as a multi-resolution analysis tool, decomposes images into low-frequency approximation coefficients and high-frequency detail coefficients through low and high-pass filters, enabling the model to simultaneously capture global structures and local details, especially key structures like microvascular networks. The multi-scale attention module establishes cross-scale spatial dependencies by reshaping and grouping the channel dimension and adopting a parallel subnetwork structure, effectively integrating contextual information while reducing computational complexity. The Adaptive Discriminator Augmentation (ADA) module monitors the training process in real-time and dynamically adjusts data augmentation intensity, preventing the discriminator from prematurely "memorizing" training samples. This allows the generator to better learn subtle features in data distribution while avoiding local optima.

3D data serves as input to the generator, which produces transformed 3D images, while the discriminator distinguishes between real and generated images. The generator and discriminator are represented as $G_{3d}$ and $D_{3d}$, respectively. $X$ represents real OCT volume data sampled from distribution $p(OCT)$, and $Y$ represents real OCTA volume data sampled from distribution $p(OCTA)$. The adversarial loss is given as:

$$\mathcal{L}_{Adv3d} = E_{\mathbf{Y}\sim p(OCTA)}\left[\log\left(D_{3d}(\mathbf{Y})\right)\right] + E_{\boldsymbol{X}\sim p(OCT)}\left[\log\left(1 - D_{3d}\left(G_{3d}(\boldsymbol{X})\right)\right)\right] \quad (1)$$

To constrain the difference between generated images and real images, and to reduce blurring, we employ L1 loss to ensure consistency between images.

$$\mathcal{L}_{\mathrm{L}13d} = \left|\left|Y - \mathrm{G}_{3d}(\mathrm{X})\right|\right|_1 \quad (2)$$

The total loss for the 3D pix2pix GAN is (we set $\lambda_A$ to 10 in our experiments):

$$\mathcal{L}_{3dGAN} = \mathcal{L}_{Adv3d} + \lambda_A \mathcal{L}_{\mathrm{L}13d} \quad (3)$$

### 3.2 Vessel Structure Matcher (VSM)

When generating OCTA images containing detailed vascular information, ensuring consistency and accuracy of vascular structures between generated images and target images is crucial. For this purpose, we developed the Vessel Structure Matcher (VSM), a module specifically designed to supervise and evaluate the matching degree of vascular regions between generated images and target images. Using VSM, we significantly enhanced the authenticity and precision of converted vascular structures, improving their clinical interpretability and reliability.

We pre-trained a model named $G_{VSM}$, which takes three-dimensional OCTA images as input and outputs pixel-level vessel segmentation results on two-dimensional projection images. We adopted the L1 loss function as a metric to evaluate the quality of vascular structures, ensuring that the generated vascular structures reflect the actual vascular distribution as accurately as possible. $x_{OCTA}$、$\hat{x}_{OCTA}$ respectively represent real and generated 3D OCTA images.

$$\mathcal{L}_{VSM} = E_{x_{OCTA},\ \hat{x}_{OCTA}}\left|\left|G_{VSM}(x_{OCTA}) - G_{VSM}(\hat{x}_{OCTA})\right|\right|_1 \quad (4)$$

### 3.3 Hierarchical Feature Calibration (HFC)

To ensure that the generated images precisely match the target images in terms of texture and details across multiple levels, we introduced the Hierarchical Feature Calibration (HFC) module. HFC focuses on precise calibration of details at different depth levels, ensuring that generated images exhibit high authenticity at each layer, improving overall image quality and readability, while enhancing the accuracy of minute vascular structures and key anatomical landmarks, thus improving the ability to identify complex retinal diseases.

We trained two models to generate different layer images: $x_{OCTA}^{proj}$ and $\hat{x}_{OCTA}^{proj}$ respectively represent the resulting projection images obtained by averaging along the z-axis from real and generated 3D OCTA images. The functions $G_{ILM-OPL}$ and $G_{OPL-BM}$ are responsible for generating projection images between the ILM (Internal Limiting Membrane) to OPL (Outer Plexiform Layer) and OPL to BM (Bruch's Membrane) projection layers based on the input 2D projection images.

To enhance image coherence and edge clarity, we introduced Total Variation Loss, effectively reducing jagged effects and unnecessary noise. This combined strategy not only precisely matches image content but also effectively maintains smooth and natural transitions of details, ensuring that generated images present higher visual quality and authenticity while preserving key features.

We use L1 loss to measure differences between images ($\lambda_B$ set to 0.25):

$$\mathcal{L}_{proj} = \left|\left|x_{OCTA}^{proj} - \hat{x}_{OCTA}^{proj}\right|\right|_1 \tag{5}$$

$$\mathcal{L}_{ILM-OPL} = \left|\left|G_{ILM-OPL}\left(x_{OCTA}^{proj}\right) - G_{ILM-OPL}\left(\hat{x}_{OCTA}^{proj}\right)\right|\right|_1 \tag{6}$$

$$\mathcal{L}_{OPL-BM} = \left|\left|G_{OPL-BM}\left(x_{OCTA}^{proj}\right) - G_{OPL-BM}\left(\hat{x}_{OCTA}^{proj}\right)\right|\right|_1 \tag{7}$$

$$\begin{aligned}\mathcal{L}_{\mathrm{tv}3d} = &\sum_{d=1}^{D-1}\sum_{h=1}^{H}\sum_{w=1}^{W}|V(d+1,h,w) - V(d,h,w)| + \\ &\sum_{d=1}^{D}\sum_{h=1}^{H-1}\sum_{w=1}^{W}|V(d,h+1,w) - V(d,h,w)| + \\ &\sum_{d=1}^{D}\sum_{h=1}^{H}\sum_{w=1}^{W-1}|V(d,h,w+1) - V(d,h,w)|\end{aligned} \tag{8}$$

$$\mathcal{L}_{HFC} = \left(\mathcal{L}_{proj} + \mathcal{L}_{ILM-OPL} + \mathcal{L}_{OPL-BM} + \mathcal{L}_{\mathrm{tv}3d}\right) * \lambda_B \tag{9}$$

Overall, we pre-trained the VSM and HFC modules, then fixed their parameters before training the complete PupiNet model. The total loss can be represented as (in our experiments, we set $\lambda_C$ to 1):

$$\mathcal{L}_{octa_loss} = \mathcal{L}_{3dGAN} + \lambda_C * \mathcal{L}_{VSM} + \mathcal{L}_{HFC} \tag{10}$$

For the OCTA to OCT conversion task, we used ResUNet as the backbone network for the generator in training, with the total loss function as follows ($\lambda_{A'}$ set to 15):

$$\mathcal{L}_{oct_loss} = \mathcal{L}_{Adv3d} + \lambda_{A'}\mathcal{L}_{\mathrm{L1}3d} \tag{11}$$

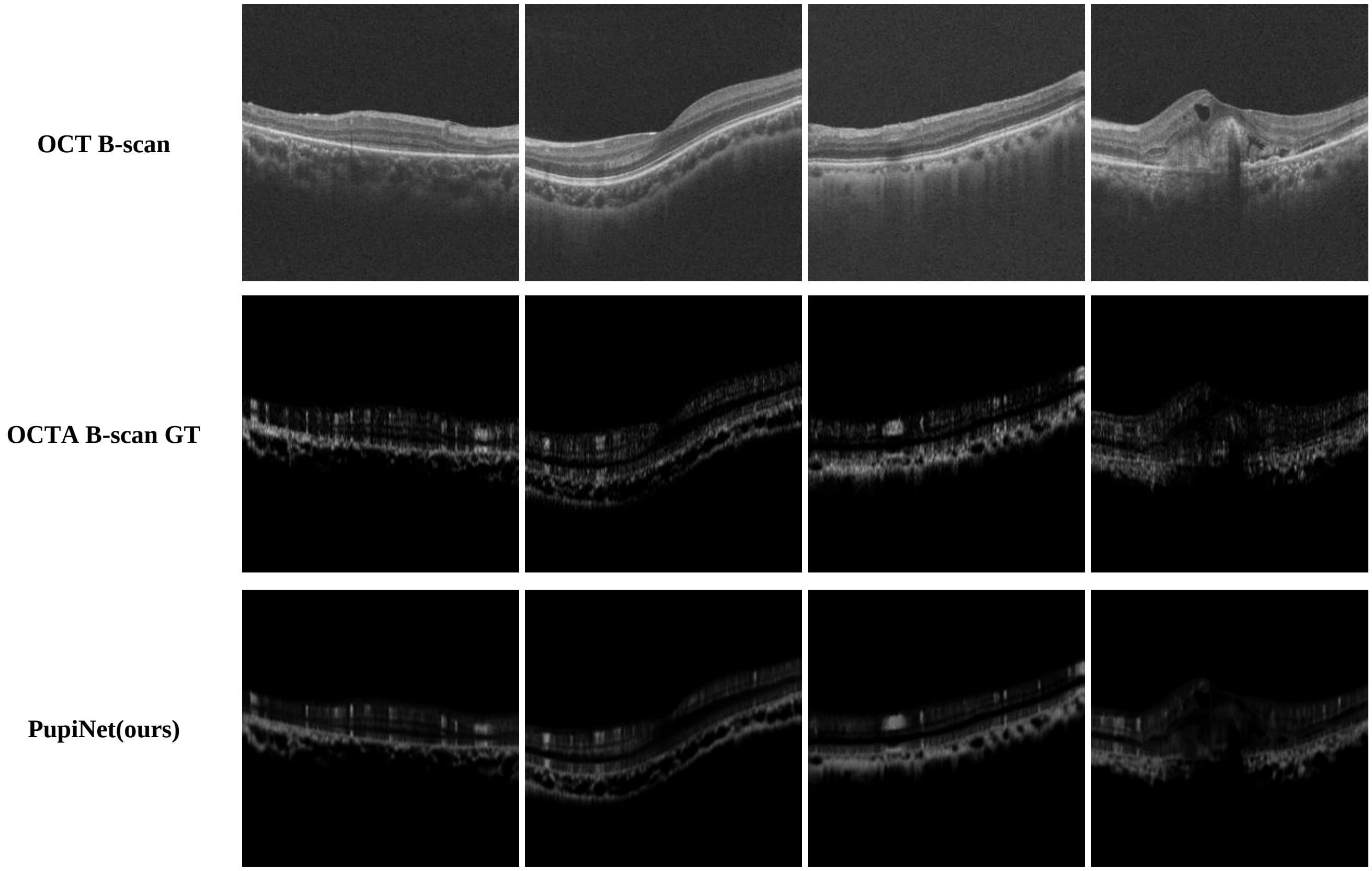

**Figure 4. Visualization of result comparisons for translated OCTA B-scan in the OCTA-6M dataset**

# 4 Experiments

## 4.1 Datasets

The dataset used in this study is derived from the publicly available OCTA-500[19], specifically the OCTA-6M subset (dataset available at https://ieee-dataport.org/open-access/octa-500).
OCTA-6M contains 300 paired OCT and OCTA data with a field of view of 6mm*6mm*2mm. The volume size is 400*400*640, with projection image dimensions of 400*400. The dataset was divided into training, validation, and test sets with sizes of 180, 20, and 100 respectively.

## 4.2 Evaluation Metrics

To evaluate the quality of the generated 3D images, this study employed standard image quality assessment metrics, specifically Mean Absolute Error (MAE), Peak Signal-to-Noise Ratio (PSNR), and Structural Similarity Index (SSIM) [20]. By computing the average values of these metrics across the entire dataset, we conducted a comprehensive analysis of the overall quality of the synthesized 3D images.

## 4.3 Experimental Details

We adopted a 3D pix2pixGAN that integrates wavelet transformation, multi-scale attention mechanisms, and adaptive discriminator augmentation strategies as the backbone network to achieve conversion from input three-dimensional volume data to the target modality. The Vessel Structure Matcher (VSM) module, based on the improved IPN-V2 model[19], is capable of generating precise vessel segmentation maps on two-dimensional projection planes from three-dimensional volume data. The Hierarchical Feature Calibration (HFC) module utilizes the UNet[21] network architecture to achieve precise conversion between projection images at different depth levels. Detailed implementation methods and parameter settings can be found in the appendix section. In Figure 4, we present a comparison among the input image, the generated image, and the real image. This illustration clearly highlights the differences among the three, providing an scientific basis for evaluating the accuracy and reality of the generated images.

## 4.4 Comparison Experiments

We conducted comparative analyses with two basic models (p2pGAN 2D, p2pGAN 3D)[15], three OCT conversion-related models (Adjacent GAN[12], 9B18CN UNet[22], TransPro[14]), and one model that has shown excellent performance in image-to-image conversion tasks (Palette[23]).

All experiments were conducted using end-to-end image generation approaches. The experimental results are presented in Table 1. "Here, '↑' indicates that higher values represent better performance, while '↓' denotes that lower values are preferable."

**Table 1: The results of the comparison with the state-of-the-arts.**

| model | PSNR ↑ | SSIM ↑ | MAE ↓ |
|---|---|---|---|
| p2pGAN 2D | 27.65 | 87.15 | 0.0995 |
| p2pGAN 3D | 29.77 | 88.18 | 0.0217 |
| Adjacent GAN | 28.05 | 85.03 | 0.1021 |
| 9B18CN UNet | 27.91 | 83.69 | 0.1135 |
| Palette | 30.02 | 87.13 | 0.0881 |
| TransPro | 30.53 | 88.35 | 0.0854 |
| **Pupi（ours）** | **30.58** | **90.64** | **0.0199** |

## 4.5 Ablation Experiments

To verify the effectiveness of the Vessel Structure Matcher (VSM) and Hierarchical Feature Calibration (HFC) modules, we conducted additional experiments by adding these two modules separately to the 3D pix2pixGAN baseline model. Table 2 presents the experimental results on the OCTA-6M dataset. Overall, the results indicate that VSM and HFC modules can effectively improve image quality, optimizing the MAE, PSNR, and SSIM metrics for OCTA images. Notably, the VSM module demonstrates superior performance by explicitly focusing on vascular morphology within 3D volumetric data, enabling comprehensive characterization of vessel continuity, branching patterns, and spatial distribution. This volumetric approach to vessel structure preservation facilitates more accurate representation of vascular networks, thereby significantly enhancing quantitative metrics.

**Table 2: Ablation studies on the VSM and HFC modules.**

| pix2pixGAN | VSM | HFC | PSNR↑ | SSIM↑ | MAE↓ |
|---|---|---|---|---|---|
| √ | | | 30.4019 | 90.30 | 0.0206 |
| √ | √ | | 30.4988 | 90.49 | **0.0199** |
| √ | | √ | 30.4476 | 90.57 | 0.0201 |
| √ | √ | √ | **30.5832** | **90.64** | **0.0199** |

When VSM and HFC modules were applied simultaneously to the baseline model, model performance improved significantly compared to using only VSM or HFC modules. Hence, we conclude that VSM and HFC modules play important roles in OCT and OCTA image conversion tasks, and their effects are additive when both modules are enabled simultaneously.
We conducted additional experiments to deeply investigate the impact of L1 loss weight in the generator on model performance, with results summarized in Table 3. The experimental results show that appropriately increasing the weight of L1 loss helps enhance the similarity between generated images and target images, demonstrating significant improvements particularly in terms of detail and edge fidelity. However, excessively high weights may lead to over-smoothing of generated images, thereby losing some necessary texture details. This indicates the important role of L1 loss in the image generation process. Selecting an appropriate L1 loss weight is one of the key factors for achieving high-fidelity image generation, as it ensures the correctness of the global structure of the image while preserving important local features and details as much as possible.

**Table 3: Ablation studies focusing on the adjustment of the $\lambda_A$ hyperparameter.**

| $\lambda_A$ | PSNR ↑ | SSIM ↑ | MAE ↓ |
|---|---|---|---|
| 100 | 30.2338 | 89.96 | 0.0206 |
| **120** | **30.4019** | **90.30** | **0.0206** |
| 130 | 30.3067 | 90.24 | 0.0205 |

To further elucidate the optimal configuration of our proposed method, we conducted ablation experiments investigating the impact of different VSM loss weights on generated image quality. As illustrated in Table 4, we evaluated $\lambda_C$ values of 3, 5, and 7, observing that $\lambda_C = 5$ yields optimal performance across all metrics. These findings suggest that a moderate VSM loss weight effectively balances the preservation of vascular morphology with overall image fidelity. When $\lambda_C$ is insufficient, the model inadequately captures fine vascular details, whereas excessive weighting potentially causes over-emphasis on vessel structures at the expense of other important image features, resulting in diminished performance. This demonstrates the critical importance of appropriately calibrating the VSM loss weight to achieve optimal vessel structure representation while maintaining high-quality overall image reconstruction.

**Table 4: Ablation studies focusing on the adjustment of the $\lambda_C$ hyperparameter when $\lambda_A$ is set to 120.**

| $\lambda_A = 120, \lambda_C$ | PSNR↑ | SSIM↑ | MAE↓ |
|---|---|---|---|
| 3 | 30.3145 | 90.38 | 0.0204 |
| **5** | **30.4988** | **90.49** | **0.0199** |
| 7 | 30.196 | 89.96 | 0.0211 |

For the conversion path from OCTA to OCT, we conducted experiments to compare the impact of L1 loss weight in the generator on the results, as presented in Table 5. The results indicate that an optimal L1 loss weight of 15 provides the best balance between structural accuracy and detail preservation, with PSNR and SSIM values higher than other tested weights.

**Table 5: Ablation studies focusing on the adjustment of the $\lambda_{A'}$ hyperparameter.**

| $\lambda_{A'}$ | PSNR↑ | SSIM↑ | MAE↓ |
|---|---|---|---|
| 5 | 26.5979 | 37.80 | 0.0679 |
| 10 | 26.4424 | 37.13 | 0.0688 |
| **15** | **27.1163** | **39.42** | **0.0652** |

## 5 Conclusion

In this paper, we innovatively proposed th PupiNet model, based on the 3D pix2pixGAN framework, to achieve efficient conversion between Optical Coherence Tomography (OCT) images and Optical Coherence Tomography Angiography (OCTA) images, and for the first time accomplished the conversion from OCTA to OCT. PupiNet combines wavelet transformation, multi-scale attention mechanisms, and adaptive discriminator augmentation strategies. By introducing the Vessel Structure Matcher (VSM) module, it ensures consistency and accuracy of vascular structures between generated images and target images, while utilizing the Hierarchical Feature Calibration (HFC) module to guarantee that textures and details in generated images match the target images at specific levels. The clinical value of our model is primarily manifested in four aspects: 1) significant reduction in healthcare expenditure through algorithmic enhancements that substitute specialized hardware investments, thereby enabling vascular imaging capabilities in primary healthcare institutions; 2) optimization of patient examination experience by generating high-quality angiographic images from a single OCT scan, thus reducing examination duration and minimizing discomfort, particularly beneficial for specialized patient populations including geriatric individuals, pediatric subjects, and patients with nystagmus; 3) enhancement of medical resource utilization efficiency by conferring dual-modality imaging functionality to conventional OCT equipment, thereby alleviating demand on advanced instrumentation while simultaneously improving early detection rates of vascular pathologies; 4) implementation of OCTA-to-OCT cross-modal conversion capabilities, providing novel data sources and validation methodologies for clinical multi-modal diagnostic protocols. Experimental results show that on the public OCTA-500 dataset, our method demonstrates superior performance and effectiveness, providing powerful support for medical image translation.

## REFERENCES

[1] Huang D, Swanson E A, Lin C P, Schuman J S, Stinson W G, Chang W. Optical Coherence Tomography[J]. 1991, 254(5035): 1178-1181.
[2] Fujimoto J, Swanson E J I O, Science V. The development, commercialization, and impact of optical coherence tomography[J]. 2016, 57(9): OCT1-OCT13.
[3] Hammes H-P, Feng Y, Pfister F, Brownlee M. Diabetic Retinopathy: Targeting Vasoregression[J]. Diabetes, 2011, 60(1): 9-16.
[4] Hong J, Tan S S, Chua J. Optical coherence tomography angiography in glaucoma[J]. Clinical and Experimental Optometry, 2024, 107(2): 110-121.
[5] Kashani A H, Chen C-L, Gahm J K, Zheng F, Richter G M, Rosenfeld P J. Optical coherence tomography angiography: a comprehensive review of current methods and clinical applications[J]. 2017, 60: 66-100.
[6] Jiang Z, Huang Z, Qiu B, Meng X, You Y, Liu X. Comparative study of deep learning models for optical coherence tomography angiography[J]. Biomed Opt Express, 2020, 11(3): 1580-1597.
[7] Ting D S W, Tan G S W, Agrawal R, Yanagi Y, Sie N M, Wong C W. Optical Coherence Tomographic Angiography in Type 2 Diabetes and Diabetic Retinopathy[J]. JAMA Ophthalmology, 2017, 135(4): 306-312.
[8] Bouma B E, De Boer J F, Huang D, Jang I-K, Yonetsu T, Leggett C L. Optical coherence tomography[J]. Nature Reviews Methods Primers, 2022, 2(1): 79.
[9] Le P, Kaur K, Patel B J S. Optical coherence tomography angiography[J]. 2024.
[10] Chua J, Tan B, Wong D, Garhöfer G, Liew X W, Popa-Cherecheanu A. Optical coherence tomography angiography of the retina and choroid in systemic diseases[J]. Progress in Retinal and Eye Research, 2024, 103: 101292.
[11] Goodfellow I, Pouget-Abadie J, Mirza M, Xu B, Warde-Farley D, Ozair S. Generative adversarial nets[J]. 2014, 27.

[12] Li P L, O'neil C, Saberi S, Sinder K, Wang K, Tan B. Deep learning algorithm for generating optical coherence tomography angiography (OCTA) maps of the retinal vasculature; proceedings of the Applications of Machine Learning 2020, F, 2020 [C]. SPIE.
[13] Pan B, Ji Z, Chen Q. MultiGAN: Multi-domain Image Translation from OCT to OCTA; proceedings of the Pattern Recognition and Computer Vision, Cham, F 2022//, 2022 [C]. Springer Nature Switzerland.
[14] Li S, Zhang D, Li X, Ou C, An L, Xu Y. Vessel-promoted OCT to OCTA image translation by heuristic contextual constraints[J]. Medical Image Analysis, 2024, 98: 103311.
[15] Isola P, Zhu J-Y, Zhou T, Efros A A. Image-to-image translation with conditional adversarial networks; proceedings of the Proceedings of the IEEE conference on computer vision and pattern recognition, F, 2017 [C].
[16] Phung H, Dao Q, Tran A. Wavelet diffusion models are fast and scalable image generators; proceedings of the Proceedings of the IEEE/CVF conference on computer vision and pattern recognition, F, 2023 [C].
[17] Ouyang D, He S, Zhang G, Luo M, Guo H, Zhan J. Efficient multi-scale attention module with cross-spatial learning; proceedings of the ICASSP 2023-2023 IEEE International Conference on Acoustics, Speech and Signal Processing (ICASSP), F, 2023 [C]. IEEE.
[18] Zhao S, Liu Z, Lin J, Zhu J-Y, Han S J a I N I P S. Differentiable augmentation for data-efficient gan training[J]. 2020, 33: 7559-7570.
[19] Secondary.Li M, Huang K, Xu Q, Yang J, Zhang Y, Ji Z. OCTA-500: A retinal dataset for optical coherence tomography angiography study[J]. Medical Image Analysis, 2024, 93: 103092.
[20] Wang Z, Bovik A C, Sheikh H R, Simoncelli E P J I T O I P. Image quality assessment: from error visibility to structural similarity[J]. 2004, 13(4): 600-612.
[21] Ronneberger O, Fischer P, Brox T. U-net: Convolutional networks for biomedical image segmentation; proceedings of the Medical image computing and computer-assisted intervention–MICCAI 2015: 18th international conference, Munich, Germany, October 5-9, 2015, proceedings, part III 18, F, 2015 [C]. Springer.
[22] Lee C S, Tyring A J, Wu Y, Xiao S, Rokem A S, Deruyter N P. Generating retinal flow maps from structural optical coherence tomography with artificial intelligence[J]. Scientific Reports, 2019, 9(1): 5694.
[23] Saharia C, Chan W, Chang H, Lee C, Ho J, Salimans T. Palette: Image-to-image diffusion models; proceedings of the ACM SIGGRAPH 2022 conference proceedings, F, 2022 [C].